\documentclass[a4paper,fleqn]{cas-dc}

\usepackage[authoryear]{natbib}

\def\tsc#1{\csdef{#1}{\textsc{\lowercase{#1}}\xspace}}
\tsc{WGM}
\tsc{QE}

\begin{document}
\let\WriteBookmarks\relax
\def\floatpagepagefraction{1}
\def\textpagefraction{.001}

\shorttitle{BetterNet: An Efficient CNN Architecture for Precise Polyp Segmentation}    

\shortauthors{O. Singh and S.S. Sengar}  

\title [mode = title]{BetterNet: An Efficient CNN Architecture with Residual Learning and Attention for Precision Polyp Segmentation}  

\author[1]{Owen Singh}[type=editor, style=chinese, orcid=0009-0005-0152-1611, linkedin=owensingh]
\ead{st20252137@outlook.cardiffmet.ac.uk}

\author[1]{Sandeep Singh Sengar}[type=editor, style=chinese, orcid=0000-0003-2171-9332, linkedin=dr-sandeep-singh-sengar-8038717b]
\ead{sssengar@cardiffmet.ac.uk}

\address[1]{School of Technologies, Cardiff Metropolitan University, Llandaff Campus, Western Avenue, Cardiff, CF5 2YB, United Kingdom}

\cormark[1]

\cortext[cor1]{Corresponding author: Dr. Sandeep Singh Sengar}

\begin{abstract}
Colorectal cancer contributes significantly to cancer-related mortality. Timely identification and elimination of polyps through colonoscopy screening is crucial in order to decrease mortality rates. Accurately detecting polyps in colonoscopy images is difficult because of the differences in characteristics such as size, shape, texture, and similarity to surrounding tissues. Current deep-learning methods often face difficulties in capturing long-range connections necessary for segmentation. This research presents BetterNet, a convolutional neural network (CNN) architecture that combines residual learning and attention methods to enhance the accuracy of polyp segmentation. The primary characteristics encompass (1) a residual decoder architecture that facilitates efficient gradient propagation and integration of multiscale features. (2) channel and spatial attention blocks within the decoder block to concentrate the learning process on the relevant areas of polyp regions. (3) Achieving state-of-the-art performance on polyp segmentation benchmarks while still ensuring computational efficiency. (4) Thorough ablation tests have been conducted to confirm the influence of architectural components. (5) The model code has been made available as open-source for further contribution. Extensive evaluations conducted on datasets such as Kvasir-SEG, CVC ClinicDB, Endoscene, EndoTect, and Kvasir-Sessile demonstrate that BetterNets outperforms current SOTA models in terms of segmentation accuracy by significant margins. The lightweight design enables real-time inference for various applications. BetterNet shows promise in integrating computer-assisted diagnosis techniques to enhance the detection of polyps and the early recognition of cancer. Link to the code: \href{https://github.com/itsOwen/BetterNet}{https://github.com/itsOwen/BetterNet}
\end{abstract}


\maketitle

\section{Introduction}\label{sec:introduction}

\begin{figure}[!htb]
    \centering
    \includegraphics[width=0.48\textwidth]{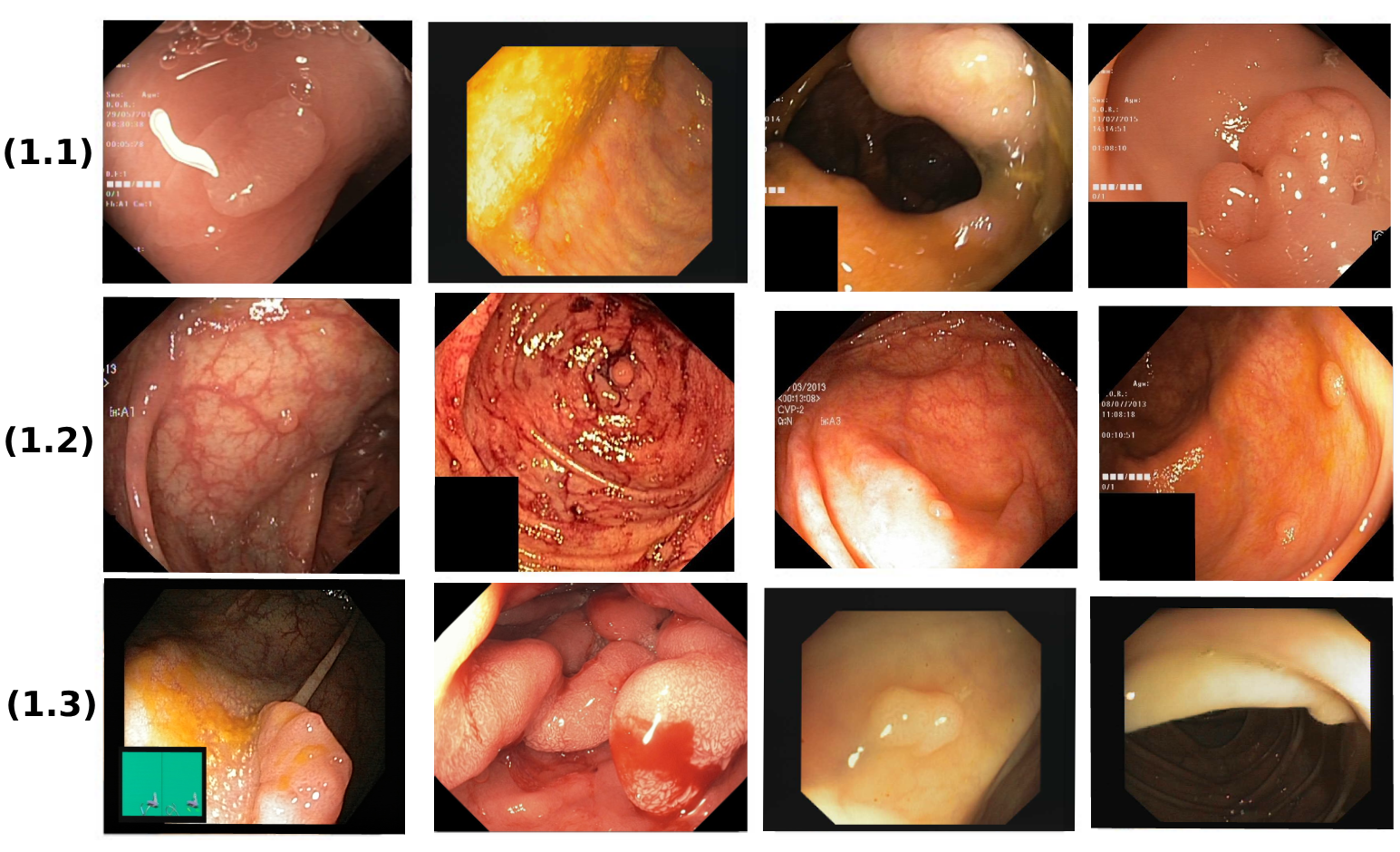}
    \caption{Various types of polyps found during colonoscopy screening: (1.1) Blurriness, low-quality images and different lighting conditions (1.2) Small size polyps, (1.3) Different shapes and sizes of polyps.}
    \label{fig:polyp_types}
\end{figure}

Colorectal cancer is a frequently occurring type of cancer and unfortunately, it is responsible for a significant number of cancer-related deaths worldwide. In 2020, there were approximately 1.93 million cases of cancer reported globally, resulting in 935,000 deaths \citep{Sung2021}. Research highlights the significance of identifying and eliminating polyps through colonoscopy screenings, which have been shown to greatly decrease both the occurrence and death rates linked to colorectal cancer \citep{Favoriti2016}. Figure \ref{fig:polyp_types} illustrates the various types of polyps that can be found during colonoscopy screenings. Nevertheless, research indicates that the detection rates for polyps during colonoscopy vary, with a range of 6\% for polyps $\geq$ 10 mm to 26\% for polyps 5 mm \citep{Kim2017}. This highlights the importance and need for implementing more precise and effective screening methods to improve results.

In an effort to help physicians identify polyps, computer-aided detection (CADe) systems use automated image analysis and pattern recognition techniques. An important component of CADe systems is the segmentation of polyps, which involves accurately outlining the boundaries of polyps in colonoscopy images, pixel by pixel. Precise segmentation maps are crucial for accurately locating, measuring, and monitoring polyps over time. However, as shown in Figure \ref{fig:polyp_types}, this task becomes challenging \citep{Liu2023} due to factors such as variations in size, shape, texture, illumination levels, and similarities with surrounding tissues. Meanwhile, current deep-learning techniques face challenges when it comes to capturing the necessary long-range connections for accurate segmentation.

This study presents BetterNet, a convolutional neural network (CNN) that has been developed for precise polyp segmentation in medical imagery. BetterNet incorporates advanced techniques to enhance polyp segmentation, resulting in improved accuracy and precision. It provides precise results and efficient calculations, making it perfect for clinical applications. Here are the main highlights:

\begin{enumerate}
    \item BetterNet introduces an innovative residual decoder architecture to improve the gradient flow and facilitate multi-scale feature fusion. This enhances the model's capacity to locate polyps in complex medical images accurately.
    
    \item BetterNet's residual decoder effortlessly integrates attention mechanisms, such as Channel Attention and Spatial Attention blocks. These strategies allow the model to focus its learning on relevant regions and contexts of polyps, leading to a significant enhancement in segmentation accuracy, particularly when dealing with small and challenging-to-detect polyps.
    
    \item BetterNet achieves a higher benchmark in polyp segmentation, showcasing outstanding performance on widely recognised benchmark datasets while also preserving computational efficiency. This achievement showcases the model's ability to seamlessly integrate into healthcare workflows, emphasising accuracy and effectiveness.
    
    \item  A comprehensive series of ablation studies was conducted to examine the impacts of crucial architectural components and training techniques used in BetterNet. These studies provide useful information on the effectiveness of design decisions, guiding subsequent advancements in polyp segmentation research.

    \item BetterNet shows a mDice improvement of approximately 3.6\% on the Kvasir-SEG \citep{Jha2020} dataset and 1.9\% on the CVC-ClincDB \citep{Bernal2015} dataset compared to CAFE-Net \citep{Liu2024}. Similarly, BetterNet demonstrates a mIoU improvement of approximately 5.1\% on the Kvasir-SEG dataset and 3.0\% on the CVC-ClinicDB dataset over CAFE-Net.
    
    \item In recognition of the importance of reproducibility and collaboration in scientific endeavors, the BetterNet model code has been made publicly available. Our purpose is to encourage transparency, foster cooperative research, and accelerate progress in examining and detecting polyps.
\end{enumerate}

The rest of the paper is organised as follows. Section \ref{sec:related-work} provides background on related works. Section \ref{sec:methodology} describes the BetterNet methodology. Section \ref{sec:training} presents training strategies. Section \ref{sec:experiments} provides comprehensive experiments and results. Section \ref{sec:discussion} analyses and discusses the results. Finally, Section \ref{sec:conclusion} concludes the paper.

\section{Related Work}\label{sec:related-work}

This section reviews the relevant literature on polyp segmentation using both traditional machine learning and recent deep learning approaches.

\subsection{Traditional Computer Vision Techniques}

Early polyp segmentation approaches relied on traditional computer vision techniques, employing hand-crafted feature extraction and rule-based classification methods to capture visual characteristics such as texture, shape, and colour~\citep{lin2024novel}. \citep{karkanis2003} made significant advancements by developing a colour wavelet-based method for detecting polyps in colonoscopy videos. They utilised linear discriminant analysis to accurately classify regions as either polyp or non-polyp. In a study conducted by \citep{Iakovidis2005}, a new method was introduced that utilised colour histograms and co-occurrence matrices. The researchers trained an SVM classifier to distinguish between polyp and non-polyp regions. The results showed that this method outperformed previous colour-based methods.

Shape-based features have also been explored, as demonstrated by the combination of colour and shape features in a study by \citep{Hwang2007, geronzi2023computer, fan2022texture}. They utilised a curvature-based shape descriptor known as shape index to analyse the local geometry of polyps. The method demonstrated high sensitivity in detecting polyps of different sizes and shapes. Nevertheless, conventional computer vision methods face certain constraints when it comes to medical image segmentation~\citep{sengar2017moving}. They fall short in capturing the complete spectrum of polyp appearances and variations, necessitate extensive parameter adjustments and domain knowledge, and encounter difficulties in learning from extensive datasets and applying that knowledge to unseen cases \citep{Liu2023}. These methods often faced challenges with specular highlights, blood vessels, and other visual artifacts common in colonoscopy images, leading to reduced segmentation accuracy and increased false positives.

\subsection{Transformer-Based Techniques}

Transformer-based models have demonstrated impressive capabilities in polyp segmentation, due to their ability to capture long-range dependencies and incorporate global contextual information \citep{Vaswani2017, zhang2024dual}. SSFormer, PVT-CASCADE, Polyp-PVT, and HSNet are examples of models that have shown the effectiveness of PVTv2 in medical image segmentation. These models have introduced different modules to improve feature fusion and extraction, as demonstrated in the study by \citep{Wang2022}. Nevertheless, these models encounter difficulties in fully harnessing encoder information, recovering intricate details, and effectively eliminating noise. Moreover, these models are very big in size and take a lot of computational power in order to provide great results.

Despite the advancements in Transformer-based approaches, such as CAFE-Net with PVT as its encoder, there is a demand for more precise and resilient polyp segmentation models to address the challenges faced. Additional research and development in this field are crucial to tackle these challenges and improve the effectiveness of polyp segmentation models.

\subsection{Deep Learning Based Techniques}

In recent years, medical image analysis has been greatly impacted by the use of deep learning techniques, specifically convolutional neural networks (CNNs)~\citep{sengar2023multi, jain2023age}. CNNs have the ability to acquire hierarchical features directly from raw image data, eliminating the necessity for manual feature extraction. There has been significant advancement in medical image segmentation tasks, particularly in polyp segmentation. \citep{Brandao2017} presented a pioneering use of CNNs for polyp segmentation. They proposed an architecture called a fully convolutional network (FCN) with an encoder-decoder structure. Their approach proved to be more effective than traditional methods when applied to a dataset of colonoscopy images.

The U-Net architecture \citep{Ronneberger2015}, which is commonly employed in medical image segmentation, has been extensively modified for polyp segmentation. The contracting and expanding paths, linked by skip connections, effectively capture contextual information and facilitate accurate localization. \citep{Akbari2018} proposed a detailed polyp segmentation approach using a modified U-Net architecture with a pre-trained encoder and a custom decoder, demonstrating excellent performance on multiple polyp segmentation datasets. In their study, \cite{Yeung2021} presented the Focus U-Net, a new deep neural network that incorporates both spatial and channel-based attention mechanisms. This unique approach, achieved through the integration of a Focus Gate module, promotes the selective learning of polyp features. The Focus U-Net incorporates several additional architectural modifications, including the inclusion of short-range skip connections and deep supervision.

Although deep learning-based approaches have demonstrated promising results in polyp segmentation, there are still various limitations and challenges that need to be tackled. Some challenges in this field involve the limited availability of comprehensive datasets, the complexity of identifying small and flat polyps, and the requirement for adaptable models that can work with various datasets and imaging conditions. In medical applications, the lack of interpretability in deep learning models is a major concern. This issue hinders the trust and adoption of these models by clinicians.

\section{BetterNet Methodology}\label{sec:methodology}

This section provides an overview of the BetterNet architecture, which consists of a pre-trained EfficientNet encoder and a residual decoder with squeeze-and-excitation blocks. The purpose of this architecture is to perform precise polyp segmentation.

\subsection{Network Architecture}

\begin{figure}[htbp]
    \centering
    \includegraphics[width=0.45\textwidth]{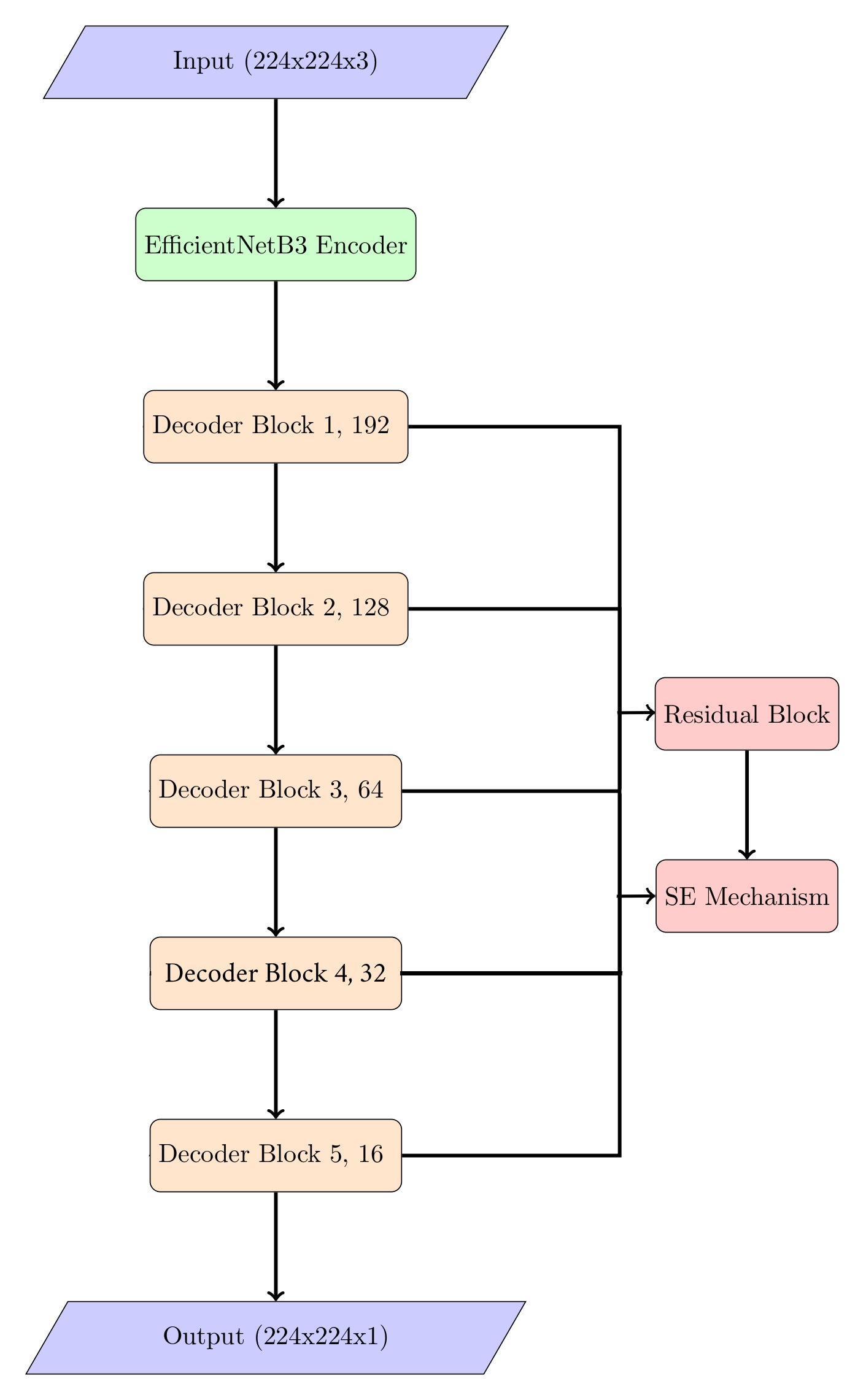}
    \caption{BetterNet Model Architecture.}
    \label{fig:betternet-block}
\end{figure}

The overall architecture of BetterNet is illustrated in Figure \ref{fig:betternet-block}. It follows a U-Net-style encoder-decoder design, which is one of the most commonly adopted techniques for semantic segmentation tasks.

The encoder module uses convolution and pooling techniques to extract structured visual information from the input frames of colonoscopies. Instead of relying on CNN layers, BetterNet uses for EfficientNet B3, as the backbone for its encoder. EfficientNets \citep{Tan2019} are mobile networks designed for performance through a careful balance of model depth, width and resolution scaling. Initializing parameters through pretraining on datasets like ImageNet results in faster convergence and better generalization.

The decoder module leverages the abstract encoder features and systematically restores spatial resolution by employing upsampling and convolutional layers. This allows for the propagation of semantic context data from the encoder to the pixel level for dense segmentation.

The decoder in BetterNet consists of a sequence of residual convolutional blocks that incorporate squeeze-and-excitation (SE) attention models. Residual connections assist in the smooth gradient flow during the training process, while the SE blocks adjust channel-wise features. Skip connections from the encoder are combined using concatenation at each stage to integrate localisation cues. Finally, a $1 \times 1$ convolution predicts the segmentation mask output.

Further details regarding the model implementation can be found in Section \ref{sec:implementation}. The following sections will cover the residual blocks and SE attention that are integrated into the BetterNet decoder.

\subsection{Residual Learning}

Residual learning, popularised in networks like ResNet \citep{He2015}, refers to adding skip connections between network layers. For a standard stack of convolutional blocks, the output $x_{l+1}$ of the $(l+1)$-th block is computed from the previous block's output $x_{l}$ as:

\[ x_{l+1} = H_{l}(x_{l}) \]

Here, $H_{l}$ denotes the layer mapping function. In contrast, residual blocks introduce an additive skip connection that bypasses the block transformations as:

\[ x_{l+1} = H_{l}(x_{l}) + x_{l} \]

This simple change offers several benefits empirically:

\begin{itemize}
    \item Enables training deeper networks by allowing gradients to flow directly through the shortcut paths.
    \item Acts as implicit feature reuse, allowing later layers to retain details from prior blocks.
    \item Regulates layer mappings to focus on residual modelling of novel information rather than directly fitting random functions.
\end{itemize}

\begin{figure}[htbp]
    \centering
    \includegraphics[width=0.3\textwidth]{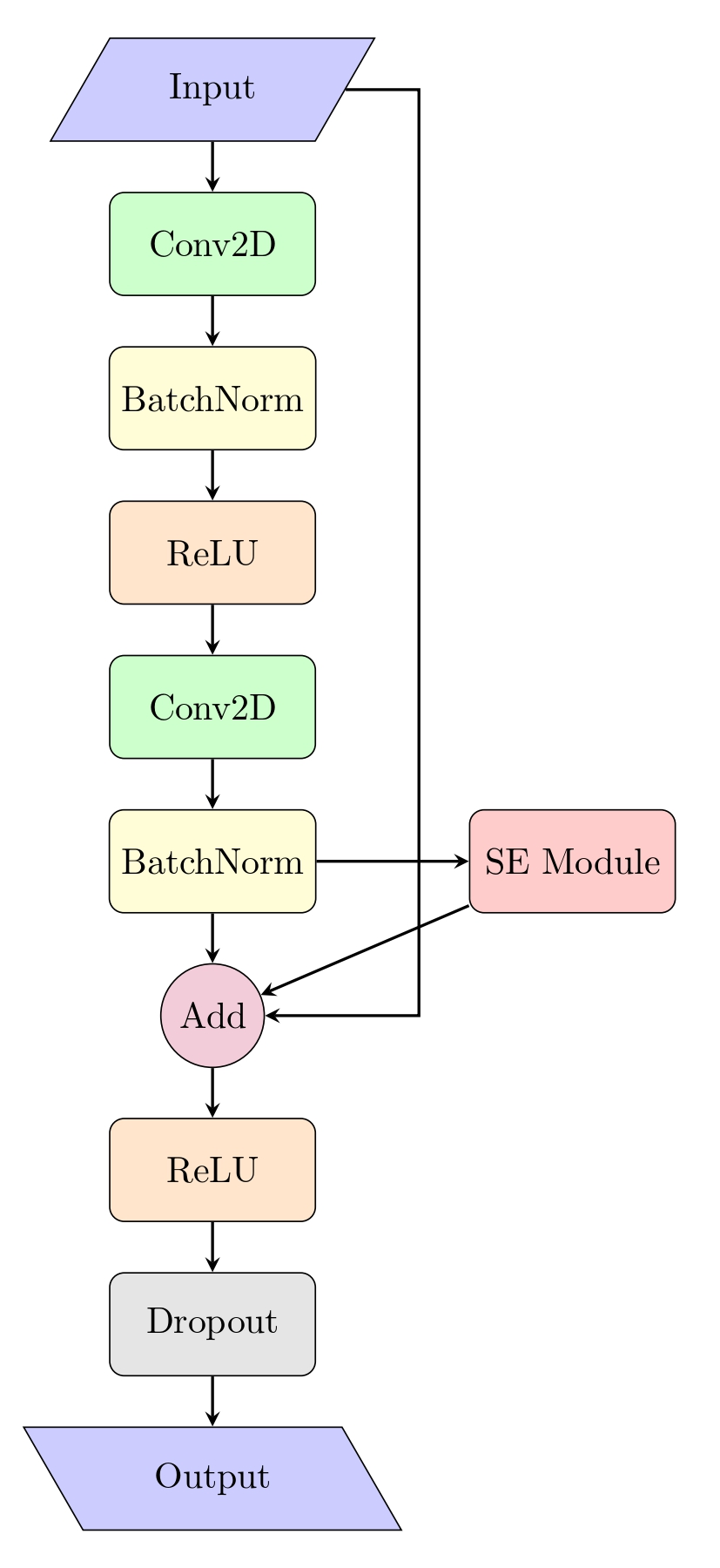}
    \caption{BetterNet residual block structure.}
    \label{fig:residual-block}
\end{figure}

In BetterNet, we adopt residual learning to aid multi-scale feature aggregation in the decoder for more effective use of hierarchical cues. The overall architecture remains lightweight without significantly increasing the model size. Figure \ref{fig:residual-block} illustrates a BetterNet residual block comprising:

\begin{enumerate}
    \item A $1 \times 1$ convolution to reduce channel dimensions.
    \item Two $3 \times 3$ convolutions for the main transformation.
    \item Squeeze-and-excitation attention modelling described next.
    \item Element-wise addition of the input shortcut path.
    \item ReLU activation for non-linearity.
\end{enumerate}

The network can improve segmentation details and maintain global localisation cues using residuals and attention.

\subsection{Squeeze-and-Excitation Attention}

Channel-wise attention mechanisms have demonstrated efficacy in recalibrating convolutional feature responses by modelling interdependencies among channels. Squeeze-and-excitation (SE) blocks are efficient components that utilise self-attention to enhance the process of feature learning.

\begin{figure}[htbp]
    \centering
    \includegraphics[width=0.22\textwidth]{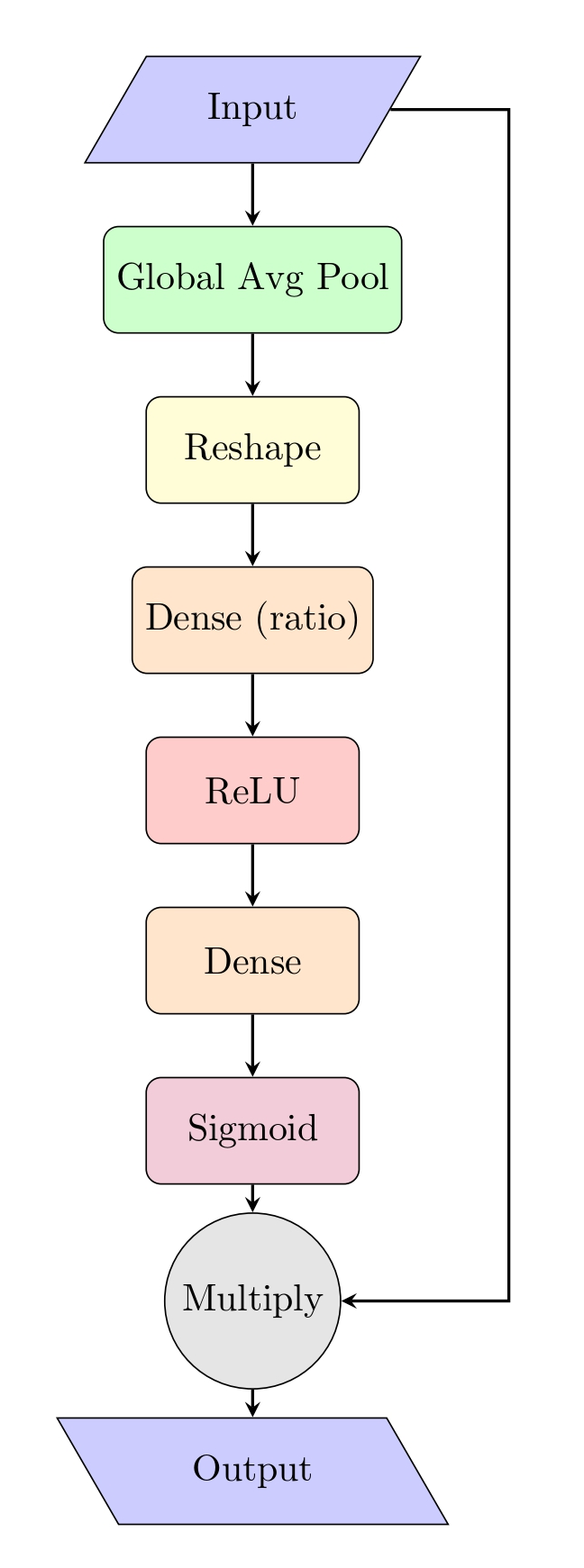}
    \caption{Squeeze-and-excitation (SE) block structure.}
    \label{fig:se-block}
\end{figure}

Figure \ref{fig:se-block} illustrates an SE block that initially consolidates spatial information into channel descriptors by using global average pooling. This squeezes the spatial dimensions while preserving channel-specific information. The descriptors are inputted into a compact MLP block to calculate scalar attention weights for each channel.

The sigmoid activation function guarantees that the weights are constrained to the range of 0 and 1. Subsequently, these weights are compounded with the input block features on an element-wise basis to accentuate significant channels while choosing diminishing irrelevant ones.

SE blocks enhance the network's feature calibration and learning capabilities by incorporating channel interdependencies into the model. The scalar reweighting incurs minimal additional computational effort. Within BetterNet, we implement SE blocks in the residual decoder units to enhance the learning of polyp features and segmentation.

\section{Training Strategies}\label{sec:training}

\subsection{Loss Function}

In order to train the BetterNet model, we utilise a combination of binary cross-entropy loss and Dice loss. The binary cross-entropy loss measures the difference between the predicted probability map and the ground truth segmentation mask. It is described as:

\[ L_{\text{BCE}} = -[y \cdot \log(p) + (1 - y) \cdot \log(1 - p)] \]

where \( y \) is the ground truth label (0 for background, 1 for polyp), and \( p \) is the predicted probability.

Dice loss is based on the Dice coefficient, which measures the overlap between the predicted segmentation and the ground truth. Dice loss is defined as:

\[ L_{\text{Dice}} = 1 - \frac{2 \cdot |P \cap G| + \epsilon}{|P| + |G| + \epsilon} \]

where \( P \) is the predicted segmentation, \( G \) is the ground truth segmentation, and \( \epsilon \) is a small constant to avoid division by zero.

The total loss is a weighted sum of binary cross-entropy loss and Dice loss:

\[ L_{\text{Total}} = \alpha \cdot L_{\text{BCE}} + (1 - \alpha) \cdot L_{\text{Dice}} \]

where \( \alpha \) is a hyperparameter that balances the contribution of the two loss terms. In our experiments, we set \( \alpha \) to 0.5.

\subsection{Optimization and Regularization}

The BetterNet model is trained using the Adam optimiser \citep{kingma2014adam}. The Adam algorithm is widely used for optimisation and it adjusts the learning rate of each parameter based on the first and second moments of the gradients. It merges the benefits of AdaGrad and RMSProp optimisers, delivering effective and consistent optimisation.

To mitigate overfitting and enhance generalisation, we utilise multiple regularisation strategies. L2 weight regularisation is employed on the convolutional layers, incorporating a penalty term into the loss function that is determined by the squared magnitude of the weights. This promotes the model to acquire more straightforward and more generalised representations.

In addition, we employ dropout regularisation in the decoder network \citep{Srivastava2014}. During training, dropout randomly deactivates some input units, effectively setting them to zero. This technique is beneficial because it reduces the tendency of neurons to become overly dependent on each other and encourages the network to develop more resilient and reliable features.

\section{Experiments and Results}\label{sec:experiments}

This section presents a comprehensive set of experiments to evaluate the proposed BetterNet model for colonoscopy polyp segmentation.

\subsection{Datasets}

Publicly available colonoscopy image datasets were used for model training and benchmarking:

We evaluate the performance of BetterNet on five widely-used polyp segmentation datasets: Kvasir-SEG \citep{Jha2020}, CVC-ClinicDB \citep{Bernal2015}, EndoTect \citep{Hicks2021}, EndoScene \cite{Vazquez2017} and Kvasir-Sessile \citep{Jha2021} (it is part of Kvasir-SEG, and extracted based on specific condition).

\begin{table}[htbp]
    \centering
    \caption{Dataset Split}
    \begin{tabular}{lcccc}
    \hline
    \textbf{Datasets} & \textbf{Number} & \textbf{Train} & \textbf{Validation} & \textbf{Test} \\
    \hline
    Kvasir-SEG & 1000 & 900 & - & 100 \\
    CVC-ClinicDB & 612 & 550 & - & 62 \\
    EndoScene & 60 & - & - & Test \\
    EndoTect & 1000 & - & - & Test \\
    Kvasir-Sessile & 196 & - & - & Test \\
    \hline
    \end{tabular}
\end{table}

\textbf{Training Datasets:}

The combined training approach, which incorporated both the Kvasir-SEG and CVC-ClinicDB datasets, has shown a marked improvement in the model's generalisability. This methodology has also provided the additional benefit of reducing the computational time that would have been required had the model been trained separately on each individual dataset. The synergistic effect of this combined training strategy has produced a more robust and efficient model, capable of adapting to a broader range of input data whilst minimising the resource expenditure associated with the training process.

\textbf{Kvasir-SEG} dataset is a comprehensive collection of 1,000 polyp images coupled with segmentation masks. The images are obtained from regular clinical colonoscopies and represent a variety of polyp sizes, shapes, and visual characteristics. The dataset is divided into 900 images for training, and 100 images for testing purposes.

\textbf{CVC-ClinicDB} dataset comprises 612 polyp images accompanied by corresponding segmentation masks. The images are acquired from 31 colonoscopy sequences, including both small and large polyps. We employ the conventional split of 550 images for training and 62 for testing.

\textbf{Generalization Test:}

\textbf{Endoscene} is one of the reputable datasets. It contains
a total number of 60 images of colorectal polyps. We have
used this dataset to test our model generalisation

\textbf{EndoTect} is a dataset that has a significant level of difficulty and consists of 1000 images of polyps. The dataset encompasses a wide range of polyp types. We have simply used this dataset to test our model's generalisation capabilities.

\textbf{Kvasir-Sessile} is a subset of the Kvasir-SEG dataset that has been refined and produced from it. This dataset consists of 196 occurrences of polyps that are less than 10 mm and are classed as either Paris class 1 sessile or Paris class IIa. We are utilising this dataset solely for testing purposes. The main goal is to highlight the subtleties of identifying smaller polyps in different clinical situations. Please refer to the section on small polyp detection for further information.

We have also used unseen images from Kvasir-SEG (10\%) and CVC-ClinicDB (10\%) for our generalisation test.

All the datasets exhibit diversity in imaging conditions and scene characteristics. For preprocessing, images were resized to 224 x 224 resolution, and pixel values were normalised to the range [0-1]. The datasets were randomly split and a combined shuffle of both training datasets was done in order to help reduce bias and improve the model's ability to generalize to unseen data.

\subsection{Implementation Details}\label{sec:implementation}

The BetterNet model was implemented using TensorFlow 2.15.0 and the Keras API \citep{Abadi2016, chollet2015keras}. The Adam optimiser was used for training, with a polynomial learning rate schedule that gradually decreased from 1e-4 to 1e-7 throughout 150 epochs. The primary objective was to reduce the binary cross-entropy loss. Validation performance was monitored by tracking measures such as the Dice coefficient, IoU, weighted-F measure, s-measure, and e-measure. Post-processing techniques, including dense conditional random fields (CRF) and morphological operations, were applied to refine the segmentation masks generated by the BetterNet model. The tests were performed on an Nvidia T4 GPU with 15GB of VRAM using a batch size of 8. We conducted training and testing of our model using Google Colab because of its convenient accessibility. Even on the free edition of Colab, our model achieved superior performance compared to the majority of state-of-the-art models. We have made the source of our model open-source for reproducibility.

\subsection{Model Configuration}

The BetterNet encoder consists of EfficientNet-B3, which has been pre-trained on ImageNet. The weights of the model are kept fixed during training. The decoder employs 64 initial filters that are subsequently doubled after each upsampling, resulting in a total of 5 stages. Decoder convolutions employ 3x3 kernels, ReLU activation, and batch normalisation \citep{Ioffe2015}. A regularisation dropout rate of 0.5 is implemented. The output sigmoid activation predicts per-pixel polyp likelihood. The model has 11.5M trainable parameters.

\subsection{Evaluation Metrics}

To evaluate the performance of our model, we use the same evaluation metrics as those used in recent studies on polyp segmentation \cite{Fan2020,Dong2021}. These metrics were calculated using the evaluation toolbox provided by the authors of \cite{Fan2020}. The toolbox computes several standard metrics for image segmentation, including mean Dice similarity coefficient (mDice) \cite{Milletari2016}, mean Intersection over Union (mIoU), mean absolute error (MAE), weighted F-measure ($F_{w}^{\beta}$) \\cite{margolin2014how}, S-measure ($S_{\alpha}$) \cite{Fan2017}, and E-measure ($E_{\xi}$) \cite{Fan2018,Fan2021}. The definitions of these metrics are as follows:

\begin{itemize}
    \item \textbf{Mean Dice coefficient (mDice)}: Measures the overlap between the predicted segmentation and ground truth masks. It is defined as:
    
    \[
    D_{\text{Dice}}(A, B) = \frac{2 \times |A \cap B|}{|A| + |B|} = \frac{2 \times TP}{2 \times TP + FP + FN}
    \]
    
    where $A$ represents the predicted segmentation, $B$ is the ground truth, $TP$ denotes true positives, $FP$ indicates false positives, and $FN$ marks false negatives.
    
    \item \textbf{Mean Intersection over Union (mIoU):} Computes the ratio of the intersection between the predicted and ground truth masks to their union. It is defined as:
    
    \[
    IOU(A, B) = \frac{A \cap B}{A \cup B} = \frac{TP}{TP + FP + FN}
    \]
    
    \item \textbf{Mean Absolute Error (MAE):} Measures the average absolute difference between the predicted and ground truth masks. It is computed as:
    
    \[
    MAE = \frac{1}{W \times H} \sum_{x=1}^{W} \sum_{y=1}^{H} |P(x, y) - Y(x, y)|
    \]
    
    where $P$ and $Y$ denote the predicted and ground truth masks, respectively, and $W$ and $H$ are the width and height of the images.
    
    \item \textbf{Weighted F-measure ($F_{\beta}^{\omega}$)}: Combines precision and recall into a single metric, with $\beta$ controlling their relative importance. It is defined as:
    
    \[
    F_{\beta}^{\omega} = \frac{(1 + \beta^2) \times Precision^{\omega} \times Recall^{\omega}}{\beta^2 \times Precision^{\omega} + Recall^{\omega}}
    \]
    
    where $Precision^{\omega}$ represents weighted precision and $Recall^{\omega}$ denotes weighted recall.
    
    \item \textbf{S-measure ($S_{\alpha}$)}: Evaluates the structural similarity between the predicted and ground truth masks. It is computed as:
    
    \[
    S_{\alpha} = \gamma \times S_{o} + (1 - \gamma) \times S_{r}
    \]
    
    where $S_{o}$ and $S_{r}$ denote the object-aware and region-aware structural similarity, respectively, and $\gamma$ is set to 0.5 by default.
    
    \item \textbf{E-measure ($E_{\xi}$)}: Captures both pixel-level matching and image-level statistics of the predicted and ground truth masks using an enhanced alignment matrix $\phi$. It is defined as:
    
    \[
    E_{\xi} = \frac{1}{W \times H} \sum_{x=1}^{W} \sum_{y=1}^{H} \phi_{FM}(x, y)
    \]
    
    where $\phi_{FM}$ represents the foreground map.
\end{itemize}

These metrics provide a comprehensive evaluation of BetterNet's polyp segmentation performance, assessing various aspects such as overlap, similarity, and structural alignment between the predicted and ground truth masks. By utilising the same evaluation toolbox as PraNet \citep{Fan2020}, Polyp-PVT \citep{Dong2021}, and CAFE-Net \citep{Liu2023}, we ensure a fair and consistent comparison of BetterNet with these state-of-the-art methods.

\subsection{Comparison with State-of-the-Art Models}

We evaluate the efficacy of BetterNet by comparing its performance to other cutting-edge polyp segmentation techniques, such as U-Net \citep{Ronneberger2015}, U-Net++ \citep{Zhou2018}, ACSNet \citep{Zhang2020}, PraNet \citep{Fan2020}, CaraNet \citep{Lou2021}, Polyp-PVT \citep{Dong2021}, HSNet \citep{Zhang2022}, SSFormer \citep{Wang2022}, PVT-CASCADE \citep{Rahman2023}, and CAFE-Net \citep{Liu2024}. These approaches encompass various architectural designs, ranging from CNN-based to transformer-based models, and have achieved competitive results on polyp segmentation benchmarks.

\begin{table}[htbp]
\centering
\caption{Performance Comparison of Different Methods on Kvasir-SEG}
\label{tab:kvasir-results}
\resizebox{\columnwidth}{!}{%
\begin{tabular}{@{}lllllllll@{}}
\toprule
\textbf{Method} & \textbf{Year} & \textbf{mDice} & \textbf{mIoU} & $\mathbf{F_{\omega\beta}}$ & $\mathbf{S_{\alpha}}$ & \textbf{mE} & \textbf{maxE} & \textbf{MAE} \\ \midrule
U-Net & 2015 & 0.806 & 0.718 & 0.784 & 0.842 & 0.890 & 0.894 & 0.052 \\
U-Net++ & 2018 & 0.797 & 0.704 & 0.767 & 0.827 & 0.884 & 0.887 & 0.056 \\
ACSNet & 2020 & 0.896 & 0.836 & 0.876 & 0.915 & 0.943 & 0.954 & 0.034 \\
PraNet & 2020 & 0.897 & 0.845 & 0.889 & 0.916 & 0.945 & 0.954 & 0.026 \\
CaraNet & 2021 & 0.911 & 0.861 & 0.902 & 0.924 & 0.953 & 0.958 & 0.024 \\
Polyp-PVT & 2021 & 0.920 & 0.871 & 0.915 & 0.927 & 0.954 & 0.962 & 0.024 \\
SSFormer & 2022 & 0.914 & 0.862 & 0.910 & 0.927 & 0.963 & 0.967 & 0.021 \\
HSNet & 2022 & 0.928 & 0.884 & 0.924 & 0.933 & \textbf{0.974} & \textbf{0.977} & 0.017 \\
PVT-CASCADE & 2023 & 0.924 & 0.875 & 0.920 & 0.929 & 0.960 & 0.964 & 0.022 \\
CAFE-Net & 2024 & 0.933 & 0.889 & 0.927 & 0.939 & 0.967 & 0.971 & 0.019 \\
\textbf{BetterNet} & 2024 & \textbf{0.969} & \textbf{0.940} & \textbf{0.972} & \textbf{0.969} & 0.968 & 0.968 & \textbf{0.006} \\ \bottomrule
\end{tabular}%
}
\end{table}

Table \ref{tab:kvasir-results} presents the quantitative results on the Kvasir-SEG dataset. BetterNet achieves the highest scores across most metrics, with an \textbf{mDice of 0.969}, \textbf{mIoU of 0.940}, \textbf{$F_{\omega\beta}$ of 0.972}, \textbf{$S_{\alpha}$ of 0.969}, and \textbf{MAE of 0.006}. It outperforms all other methods in these metrics, demonstrating its effectiveness in accurately segmenting polyps. However, HSNet surpasses BetterNet in terms of \textbf{mE (0.974)} and \textbf{maxE (0.977)}, indicating its strong performance in capturing fine details of polyps.

CAFE-Net, the previous state-of-the-art, achieves a mDice of 0.933, mIoU of 0.889, $F_{\omega\beta}$ of 0.927, $S_{\alpha}$ of 0.939, mE of 0.967, maxE of 0.971, and MAE of 0.019. BetterNet surpasses the overall performance of larger models like CAFE-Net and PVT-CASCADE, which have over 35 million parameters each while having only 11.5 million parameters itself. This highlights the efficiency and effectiveness of BetterNet's architecture design in achieving superior results across most evaluation measures on the Kvasir-SEG dataset, with HSNet showcasing its strength in capturing fine-grained details.

\begin{figure}[htbp]
\centering
\includegraphics[width=0.45\textwidth]{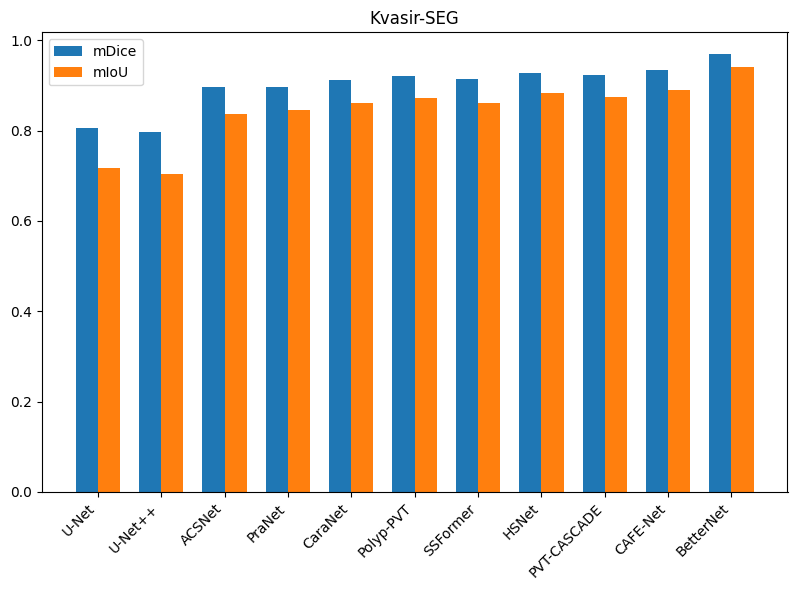}
\caption{Performance comparison of BetterNet and other state-of-the-art methods on the Kvasir-SEG dataset using the mDice and mIoU metrics.}
\label{fig:kvasir-seg-plot}
\end{figure}

Figure \ref{fig:kvasir-seg-plot} illustrates the performance comparison of BetterNet and other state-of-the-art methods on the Kvasir-SEG dataset using the mDice and mIoU metrics. The bar plot clearly shows that BetterNet outperforms all other models, achieving the highest mDice and mIoU scores. This visual representation highlights the superiority of BetterNet in accurately segmenting polyps compared to existing methods on the Kvasir-SEG dataset.

\begin{table}[htbp]
  \centering
  \caption{Performance Comparison of Different Methods on CVC-ClinicDB}
  \label{tab:clinicdb-results}
  \resizebox{\columnwidth}{!}{%
  \begin{tabular}{@{}lllllllll@{}}
    \toprule
    \textbf{Method} & \textbf{Year} & \textbf{mDice} & \textbf{mIoU} & $\mathbf{F_{\omega\beta}}
    $ & $\mathbf{S_{\alpha}}$ & \textbf{mE} & \textbf{maxE} & \textbf{MAE} \\ \midrule
    U-Net           & 2015           & 0.830          & 0.769         & 0.829                       & 0.887                 & 0.918       & 0.921         & 0.020        \\
    U-Net++         & 2018           & 0.838          & 0.777         & 0.837                       & 0.895                 & 0.920       & 0.926         & 0.019        \\
    ACSNet          & 2020           & 0.927          & 0.872         & 0.920                       & 0.950                 & 0.979       & 0.987         & 0.009        \\
    PraNet          & 2020           & 0.916          & 0.870         & 0.915                       & 0.946                 & 0.973       & 0.978         & 0.007        \\
    CaraNet         & 2021           & 0.936          & 0.887         & 0.934                       & 0.953                 & 0.985       & 0.990         & 0.006        \\
    Polyp-PVT       & 2021           & 0.939          & 0.892         & 0.939                       & 0.948                 & 0.987       & 0.990         & 0.006        \\
    SSFormer        & 2022           & 0.908          & 0.861         & 0.913                       & 0.938                 & 0.965       & 0.980         & 0.009        \\
    HSNet           & 2022           & 0.936          & 0.890         & 0.934                       & 0.949                 & 0.982       & 0.986         & 0.006        \\
    PVT-CASCADE     & 2023           & 0.940          & 0.895         & 0.940                       & 0.952                 & 0.983       & 0.987         & 0.006        \\
    CAFE-Net        & 2024           & 0.943          & 0.899         & 0.941                       & 0.957                 & \textbf{0.986}       & \textbf{0.991}         & 0.006        \\
    \textbf{BetterNet}       & 2024           & \textbf{0.962}         & \textbf{0.929}        & \textbf{0.964}                      & \textbf{0.963}                & 0.962      & 0.962        & \textbf{0.005}       \\ \bottomrule
  \end{tabular}%
  }
\end{table}

Table \ref{tab:clinicdb-results} showcases the performance comparison of different methods on the CVC-ClinicDB dataset. BetterNet continues to demonstrate its superiority by achieving the highest scores in several key metrics. It obtains an \textbf{mDice of 0.962}, \textbf{mIoU of 0.929}, \textbf{$F_{\omega\beta}$ of 0.964}, \textbf{$S_{\alpha}$ of 0.963}, and \textbf{MAE of 0.005}. These remarkable results highlight BetterNet's exceptional ability to accurately segment polyps in challenging clinical scenarios, surpassing the performance of other state-of-the-art methods.

It is worth noting that CAFE-Net, the previous state-of-the-art, also exhibits strong performance in certain evaluation measures. Specifically, CAFE-Net achieves the highest scores in \textbf{mE (0.986)} and \textbf{maxE (0.991)}, indicating its effectiveness in capturing fine-grained details of polyps.

BetterNet's outstanding performance on the CVC-ClinicDB dataset further validates its robustness and generalization capability. The model's ability to consistently achieve high scores across multiple evaluation measures demonstrates its potential for reliable and accurate polyp segmentation in real-world clinical applications. Moreover, BetterNet's superior performance is achieved with a significantly lower parameter count compared to larger models, highlighting its efficiency and practicality.

\begin{figure}[htbp]
\centering
\includegraphics[width=0.45\textwidth]{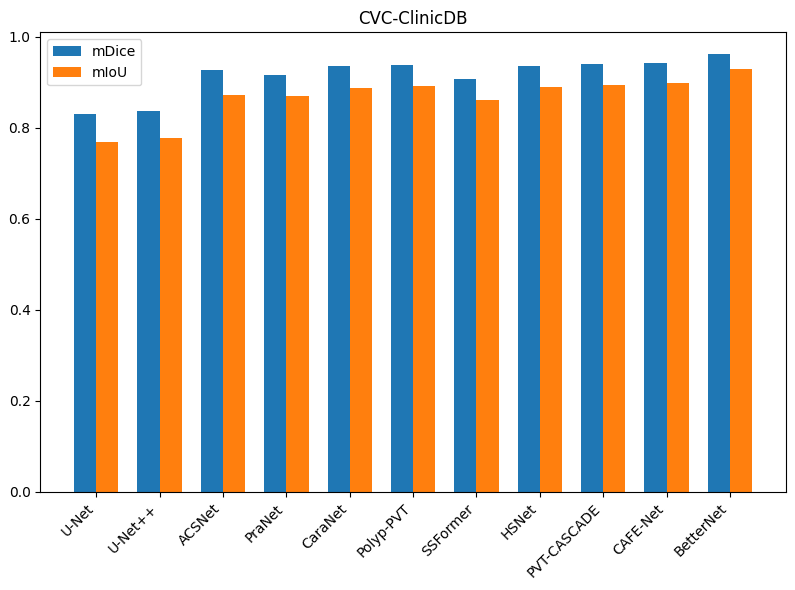}
\caption{Performance comparison of BetterNet and other state-of-the-art methods on the CVC-ClinicDB dataset using the mDice and mIoU metrics.}
\label{fig:cvc-clinicdb-plot}
\end{figure}

Similarly, Figure \ref{fig:cvc-clinicdb-plot} presents the performance comparison on the CVC-ClinicDB dataset. BetterNet demonstrates its exceptional performance by surpassing all other state-of-the-art methods in terms of mDice and mIoU metrics. The bar plot provides a clear visual indication of BetterNet's superior ability to accurately segment polyps in the CVC-ClinicDB dataset compared to other methods.

\begin{table}[htbp]

\centering

\caption{Generalisation test on EndoScene Dataset}

\label{tab:endoscene-results}

\resizebox{\columnwidth}{!}{%

\begin{tabular}{@{}lllllllll@{}}

\toprule

\textbf{Method} & \textbf{Year} & \textbf{mDice} & \textbf{mIoU} & $\mathbf{F_{\omega\beta}}$ & $\mathbf{S_{\alpha}}$ & \textbf{mE} & \textbf{maxE} & \textbf{MAE} \\ \midrule

U-Net & 2015 & 0.681 & 0.593 & 0.647 & 0.895 & 0.852 & 0.858 & 0.018 \\

U-Net++ & 2018 & 0.670 & 0.589 & 0.647 & 0.873 & 0.818 & 0.840 & 0.021 \\

ACSNet & 2020 & 0.819 & 0.725 & 0.762 & 0.893 & 0.908 & 0.976 & 0.015 \\

PraNet & 2020 & 0.869 & 0.791 & 0.831 & 0.939 & 0.940 & 0.951 & 0.011 \\

CaraNet & 2021 & 0.900 & 0.833 & 0.879 & 0.939 & 0.967 & 0.980 & 0.006 \\

Polyp-PVT & 2021 & 0.864 & 0.793 & 0.851 & 0.916 & 0.950 & 0.954 & 0.007 \\

SSFormer & 2022 & 0.882 & 0.811 & 0.858 & 0.924 & 0.954 & 0.961 & 0.007 \\

HSNet & 2022 & 0.884 & 0.807 & 0.862 & 0.937 & 0.960 & 0.961 & 0.008 \\

PVT-CASCADE & 2023 & 0.893 & 0.831 & 0.872 & 0.929 & 0.945 & 0.957 & 0.010 \\

CAFE-Net & 2024 & \textbf{0.901} & \textbf{0.834} & 0.882 & \textbf{0.939} & \textbf{0.971} & \textbf{0.981} & 0.006 \\

\textbf{BetterNet} & 2024 & 0.896 & 0.824 & \textbf{0.910} & 0.906 & 0.896 & 0.896 & \textbf{0.006} \\ \bottomrule

\end{tabular}%
}
\end{table}

Table \ref{tab:endoscene-results} presents the generalization test results of different methods on the EndoScene dataset. Although BetterNet does not achieve the highest scores in all metrics, it demonstrates competitive performance while maintaining a low parameter count and fast inference speed.

BetterNet obtains an \textbf{$F_{\omega\beta}$ of 0.910}, surpassing all other methods in this key metric. This highlights BetterNet's ability to effectively balance precision and recall in the segmentation task. Additionally, BetterNet achieves an \textbf{MAE of 0.006}, matching the lowest error rate among all the compared methods.

It is important to note that while BetterNet's performance in some metrics is slightly lower than the state-of-the-art CAFE-Net, the difference is marginal. CAFE-Net achieves the highest scores in \textbf{mDice (0.901)}, \textbf{mIoU (0.834)}, \textbf{$S_{\alpha}$ (0.939)}, \textbf{mE (0.971)}, and \textbf{maxE (0.981)}. However, BetterNet's scores in these metrics are very close, demonstrating its competitiveness.

The strength of BetterNet lies in its ability to maintain a good balance between performance and efficiency. Despite having a lower parameter count of 11 million, BetterNet achieves results that are comparable to the state-of-the-art methods. This makes BetterNet an attractive choice for real-time applications or scenarios where computational resources are limited.

\begin{figure}[htbp]
\centering
\includegraphics[width=0.45\textwidth]{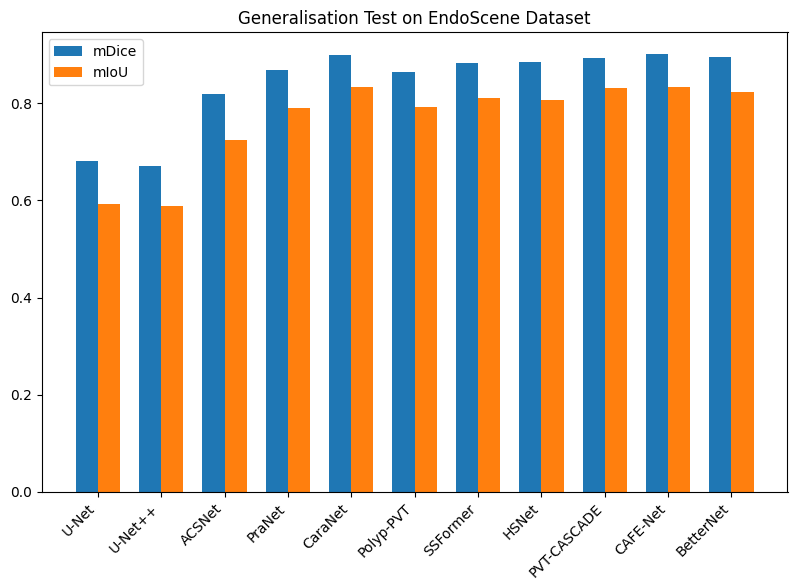}
\caption{Performance comparison of BetterNet and other state-of-the-art methods on the EndoScene dataset using the mDice and mIoU metrics.}
\label{fig:endoscene-plot}
\end{figure}

Figure \ref{fig:endoscene-plot} shows the analysis of various state-of-the-art methods on the EndoScene dataset. The bar graphs depict the performance of each method in terms of mDice and mIoU scores. On the EndoScene dataset, CAFE-Net achieves the highest mDice score of 0.901, closely followed by BetterNet at 0.896. Similarly, CAFE-Net outperforms other methods on the mIoU metric, attaining a score of 0.834. The proposed BetterNet demonstrates competitive performance, securing the second-highest mIoU score of 0.824.

\begin{table}[htbp]
\centering
\caption{Generalisation test on EndoTect Dataset}
\label{tab:endotect-results}
\resizebox{\columnwidth}{!}{%
\begin{tabular}{@{}lllllllll@{}}
    \toprule
    \textbf{Method} & \textbf{Year} & \textbf{mDice} & \textbf{mIoU} & $\mathbf{F_{\omega\beta}}$ & $\mathbf{S_{\alpha}}$ & \textbf{mE} & \textbf{maxE} & \textbf{MAE} \\ \midrule
    BetterNet & 2024 & 0.886 & 0.818 & 0.897 & 0.902 & 0.886 & 0.886 & 0.030 \\ \bottomrule
    \end{tabular}%
    }
\end{table}

Table \ref{tab:endotect-results} presents the results on the EndoTect dataset, which is used to assess the generalisation ability of the models. BetterNet achieves a mDice score of 0.886, demonstrating its strong performance on this unseen dataset. The high $F_{\omega\beta}$ score of 0.897 indicates BetterNet's ability to balance precision and recall in polyp segmentation effectively. Furthermore, the mIoU and $S_{\alpha}$ scores of 0.8185 and 0.902, respectively, showcase BetterNet's robustness in handling diverse polyp types and challenging scenarios. These results highlight BetterNet's generalisation capabilities and potential to perform well on various polyp segmentation tasks.

\begin{table}[htbp]
\centering
\caption{Generalisation test on Kvasir-Sessile Dataset}
\label{tab:kvasir-sessile-results}
\resizebox{\columnwidth}{!}{%
\begin{tabular}{@{}lllllllll@{}}
    \toprule
    \textbf{Method} & \textbf{Year} & \textbf{mDice} & \textbf{mIoU} & $\mathbf{F_{\omega\beta}}$ & $\mathbf{S_{\alpha}}$ & \textbf{mE} & \textbf{maxE} & \textbf{MAE} \\ \midrule
    BetterNet & 2024 & 0.936 & 0.889 & 0.938 & 0.941 & 0.936 & 0.936 & 0.008 \\ \bottomrule
    \end{tabular}%
    }
\end{table}

Table \ref{tab:kvasir-sessile-results} showcases the performance metrics evaluated on the Kvasir-Sessile dataset, serving as a testament to the model's generalisation proficiency. With a mIoU of 0.889 and a mDice score of 0.936, BetterNet demonstrates commendable segmentation accuracy. Notably, it attains the highest $F_{\omega\beta}$ score of 0.938, indicating a balanced trade-off between precision and recall. Moreover, the $S_{\alpha}$ score of 0.941 highlights the model's adeptness in handling diverse polyp types and challenging scenarios. Overall, these results underscore BetterNet's robustness and efficacy in polyp detection across varying clinical contexts.

The results showcase the effectiveness of the BetterNet model in precisely segmenting polyps across various datasets. By including an efficient encoder-decoder architecture, residual learning, attention mechanisms, and skip connections, BetterNet is able to capture vital information and provide accurate segmentation maps effectively. Despite its lightweight design with only 11.5 million parameters, BetterNet consistently outperforms larger and more complex models, highlighting its efficiency and potential for real-time clinical applications.

\subsection{Ablation Studies}

To investigate the impact of different components in the BetterNet model, we conducted ablation studies on the Kvasir-SEG dataset and tested it on Kvasir-Sessile. We evaluated the performance of three variants of BetterNet:

\begin{enumerate}
\item BetterNet \textbf{w/o Attention}: This variant removes the channel attention and spatial attention modules from the BetterNet model.
\item BetterNet \textbf{w/o Skip Connections}: This variant removes the skip connections between the encoder and decoder in the BetterNet model.
\item BetterNet \textbf{w/o Pre-training}: This variant initializes the EfficientNet-B3 backbone randomly instead of using pre-trained weights.
\end{enumerate}

\begin{table}[htbp]
\centering
\caption{Ablation studies test on the Kvasir-Sessile dataset.}
\label{tab:ablation-results}
\resizebox{\columnwidth}{!}{%
\begin{tabular}{|l|c|c|c|}
\hline
\textbf{Method} & \textbf{mIoU} & \textbf{mDice} & $\mathbf{F_{\omega\beta}}$ \\ \hline
BetterNet & 0.889 & 0.936 & 0.938 \\ \hline
BetterNet w/o Attention & 0.502 & 0.632 & 0.673 \\ \hline
BetterNet w/o Skip Connections & 0.480 & 0.611 & 0.645 \\ \hline
BetterNet w/o Pre-training & 0.507 & 0.631 & 0.686 \\ \hline
\end{tabular}%
}
\end{table}

Table \ref{tab:ablation-results} presents the results of the ablation studies. Removal of attention mechanisms (BetterNet \textbf{w/o Attention}) led to significant decreases in mIoU, mDice and $\mathbf{F_{\omega\beta}}$ scores (mIoU: \textbf{0.889} to \textbf{0.502}, mDice: \textbf{0.936} to \textbf{0.632}, $\mathbf{F_{\omega\beta}}$: \textbf{0.938} to \textbf{0.673}), indicating a decline in segmentation accuracy. This suggests attention mechanisms play a crucial role in capturing relevant features and improving segmentation performance.

Similarly, the absence of skip connections (BetterNet \textbf{w/o Skip Connections}) resulted in notable performance drops across all metrics (mIoU: \textbf{0.889} to \textbf{0.480}, mDice: \textbf{0.936} to \textbf{0.611}, $\mathbf{F_{\omega\beta}}$: \textbf{0.938} to \textbf{0.645}). This underscores the importance of skip connections in preserving spatial information and facilitating accurate segmentation.

When pre-trained weights were not used (BetterNet \textbf{w/o Pre-training}), there was also a decrease in performance across all metrics (mIoU: \textbf{0.889} to \textbf{0.507}, mDice: \textbf{0.936} to \textbf{0.631}, $\mathbf{F_{\omega\beta}}$: \textbf{0.938} to \textbf{0.686}). This suggests that initialising the model with pre-trained weights is beneficial for improving segmentation accuracy.

\begin{figure}[htbp]
\centering
\includegraphics[width=0.45\textwidth]{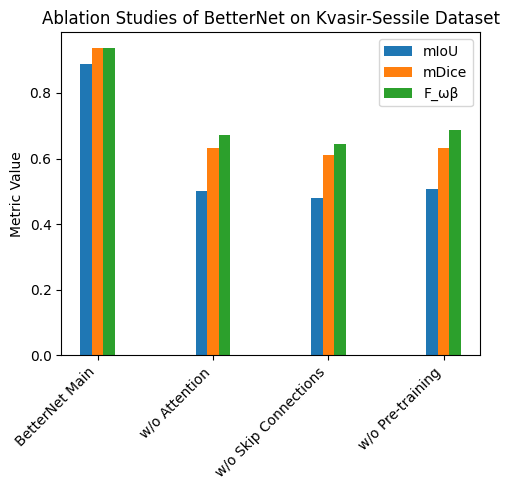}
\caption{Ablation study results on the Kvasir-Sessile dataset, comparing the performance of BetterNet and its variants without attention, skip connections, and pre-training.}
\label{fig:ablation-plot}
\end{figure}

Figure \ref{fig:ablation-plot} provides a visual representation of the ablation study results. The bar plot clearly demonstrates the impact of each component on the segmentation performance. The full BetterNet model achieves the highest scores across all metrics, while the variants without attention, skip connections, and pre-training show significant performance degradation. This visual analysis reinforces the importance of these components in the BetterNet architecture for accurate polyp segmentation.

It's important to note that we retrained the model on the Kvasir-SEG dataset for each variant and then evaluated its performance on the Kvasir-Sessile dataset, which contains polyps smaller than 10mm. This ensures that the results are consistent and applicable to real-world scenarios involving polyp segmentation tasks.

These ablation studies validate the design choices made in the BetterNet model and demonstrate the individual contributions of attention mechanisms, skip connections, and pre-training to the overall segmentation performance.

\subsection{Qualitative Results}

\begin{figure}[htbp]
    \centering
    \includegraphics[width=0.5\textwidth]{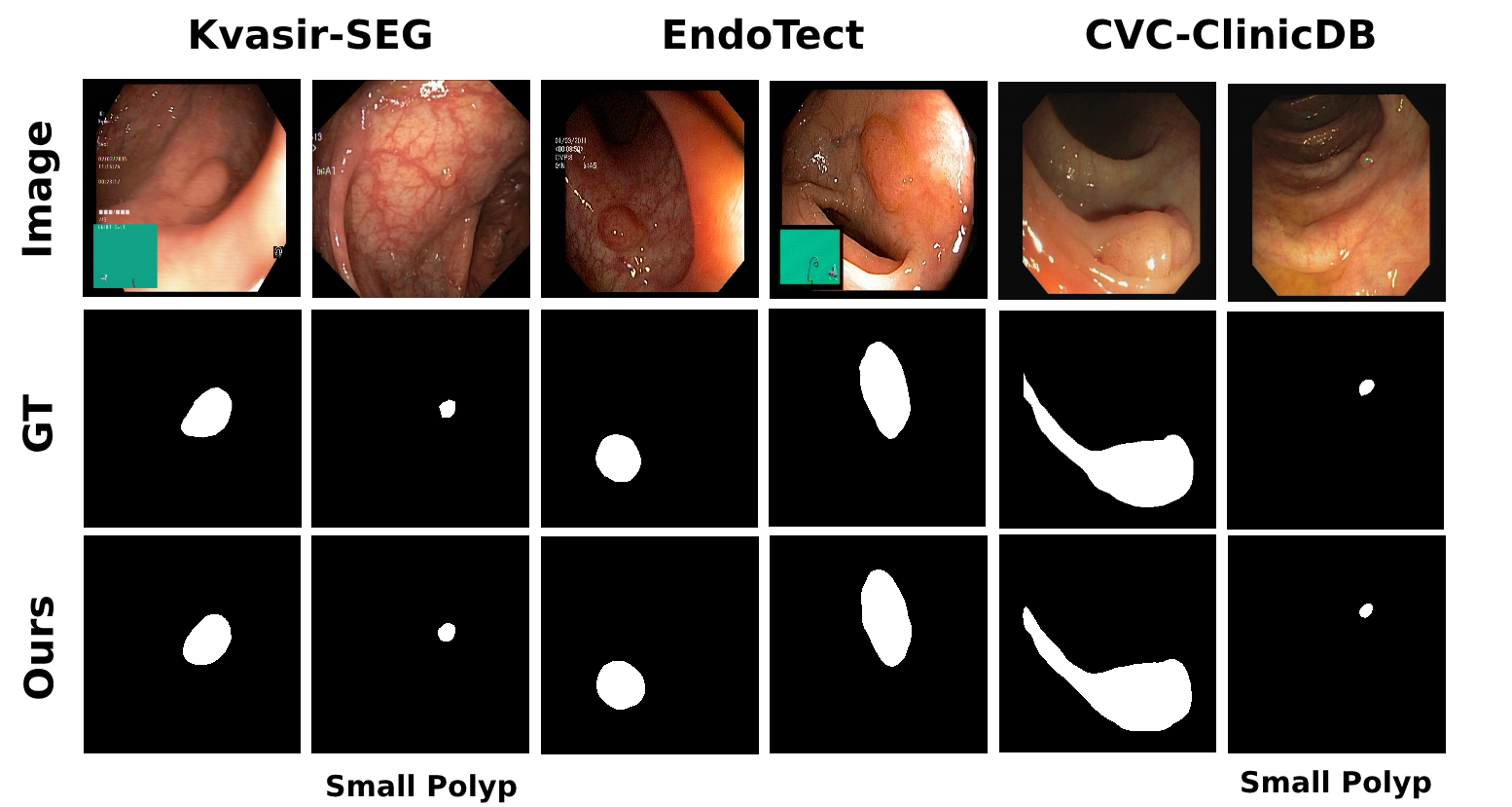}
    \caption{Qualitative results of BetterNet on sample images.}
    \label{fig:qualitative-results}
\end{figure}

Figure \ref{fig:qualitative-results} presents the qualitative results of BetterNet on sample images from the Kvasir-SEG, CVC-ClinicDB, EndoTect and Kvasir-Sessile datasets. The segmentation maps generated by BetterNet closely match the ground truth annotations, accurately delineating the boundaries of the polyps. The model can handle polyps of various sizes, shapes, and appearances, demonstrating its robustness and generalisability.

\subsection{Small Polyp Segmentation}

\begin{figure}[htbp]
\centering
\includegraphics[width=0.5\textwidth]{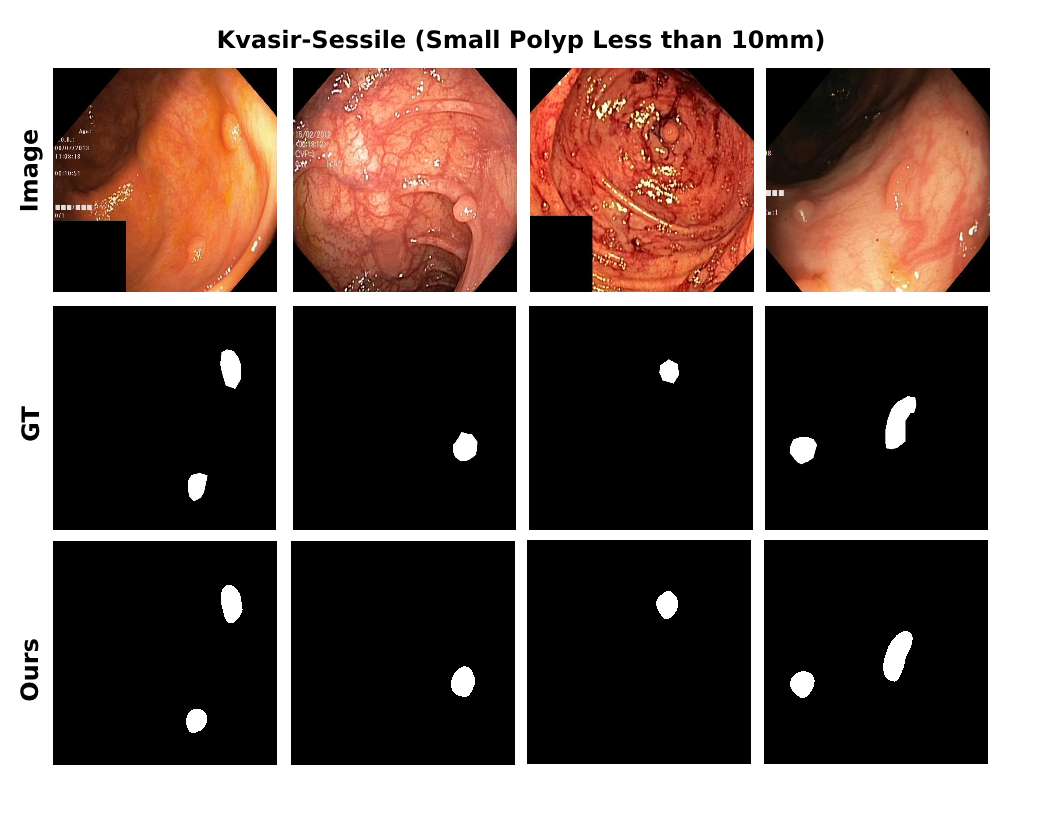}
\caption{Qualitative results of BetterNet on Kvasir-Sessile Polyps Smaller than 10mm.}
\label{fig:small-polyp-results}
\end{figure}

Accurate detection of small polyps is crucial in colonoscopies, as they are more likely to be missed during the procedure. To evaluate BetterNet's performance in such challenging cases, we tested it on the Kvasir-Sessile dataset, which consists of polyps smaller than 10mm in size. Figure \ref{fig:small-polyp-results} illustrates BetterNet's ability to segment these small polyps effectively.

BetterNet showcases its superiority in segmenting challenging small polyps across all datasets. On the Kvasir-Sessile subset of Kvasir-SEG, BetterNet achieves a mIoU of 0.8563 without being specifically trained on this dataset. Notably, BetterNet outperforms other models, such as ResUNet++ with TTA \citep{Jha2021}, which achieves a mIoU of 0.6606. The visual results in Figure \ref{fig:small-polyp-results} further demonstrate the model's precise detection of low-size polyps. These findings suggest that BetterNet has the potential to significantly improve the detection of precancerous lesions in clinical practice.

\subsection{Computational Efficiency}

We evaluate the computational efficiency of BetterNet in terms of inference time, number of parameters, and floating-point operations (GFLOPs). Table \ref{tab:efficiency-comparison} compares these metrics for BetterNet with other state-of-the-art methods.

\begin{table}[htbp]

\centering

\caption{Computational efficiency comparison}

\label{tab:efficiency-comparison}

\resizebox{\columnwidth}{!}{%

\begin{tabular}{|l|c|c|c|c|}

\hline

\textbf{Method} & \textbf{Year} & \textbf{Type} & \textbf{Params (M)} & \textbf{GFLOPs (G)} \\ \hline

U-Net & 2015 & CNN & 31.044 & 103.489 \\ \hline

U-Net++ & 2018 & CNN & 47.176 & 377.449 \\ \hline

ACSNet & 2020 & CNN & 29.451 & 17.978 \\ \hline

PraNet & 2020 & CNN & 30.498 & 13.150 \\ \hline

CaraNet & 2021 & CNN & 44.594 & 21.749 \\ \hline

Polyp-PVT & 2021 & Transformer & 25.108 & 10.018 \\ \hline

SSFormer & 2022 & Transformer & 29.310 & 19.112 \\ \hline

HSNet & 2022 & Transformer & 29.257 & 10.943 \\ \hline

PVT-CASCADE & 2023 & Transformer & 35.273 & 15.404 \\ \hline

CAFE-Net & 2024 & Transformer & 35.530 & 16.119 \\ \hline

\textbf{BetterNet} & 2024 & CNN & \textbf{11.499} & \textbf{0.0230} \\ \hline

\end{tabular}%

}

\end{table}

BetterNet successfully strikes a desirable equilibrium between segmentation performance and computational efficiency. The model has a mean inference time of 179.2 milliseconds per image, indicating its suitability for real-time applications. Although the inference timings of other techniques are not specified, BetterNet's efficient architecture and lightweight design contribute to its fast inference speed.

\begin{figure}[htbp]
\centering
\includegraphics[width=0.45\textwidth]{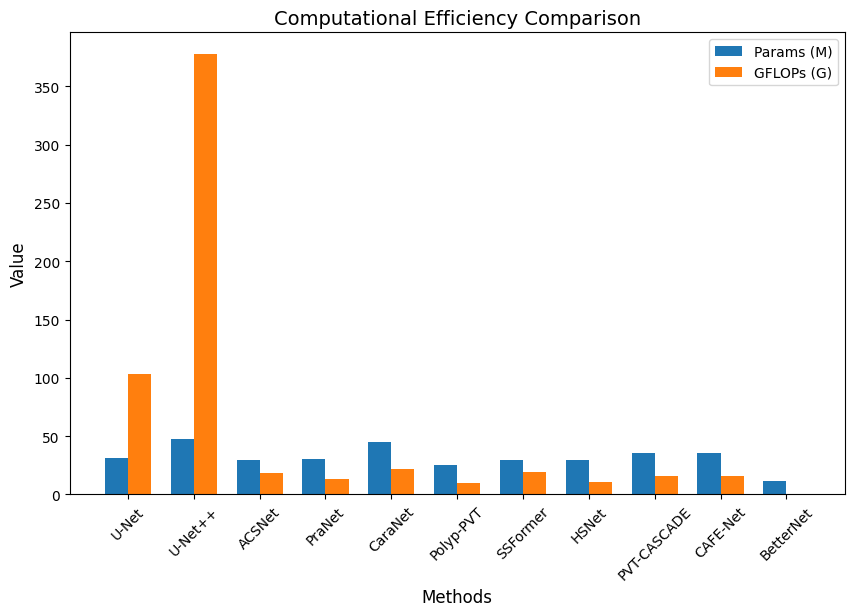}
\caption{Computational Efficiency}
\label{fig:efficient-plot}
\end{figure}

Figure \ref{fig:efficient-plot} provides a visual comparison of the computational efficiency of BetterNet and other state-of-the-art methods. The bar plot clearly illustrates the significant reduction in model parameters and GFLOPs achieved by BetterNet compared to its counterparts.

When it comes to model size, BetterNet has 11.5 million parameters, which is substantially smaller than most CNN-based models such as U-Net (31.044M), U-Net++ (47.176M), ACSNet (29.451M), PraNet (30.498M), and CaraNet (44.594M). The reduced model size is advantageous for deployment on devices with limited resources and decreases the memory requirements for both training and inference.

Furthermore, BetterNet has an approximate computational power of 0.0230 GFLOPs, as determined by its model size. This value is remarkably lower than all other models being compared, as evident from Figure \ref{fig:efficient-plot}. With a low GFLOP count, BetterNet demonstrates efficient performance in polyp segmentation, requiring minimal computational resources.

In summary, BetterNet achieves state-of-the-art segmentation performance while maintaining a high level of computational efficiency. The fast inference time, small model size, and low GFLOP count of this approach make it a viable option for real-time polyp segmentation in resource-constrained scenarios, such as clinical applications and embedded devices.

\section{Discussion}\label{sec:discussion}

This section provides further analysis and insights based on the experimental results.

\subsection{Key Observations}

\begin{enumerate}
    \item BetterNet demonstrates state-of-the-art segmentation performance on many benchmark datasets, validating the advantages of including residuals and attention. Continuous improvements in both the evaluation metrics and datasets demonstrate the strength and resilience of the system.

    \item By performing ablations, it has been discovered that the joint modelling of residuals and SE attention has a complementing and beneficial effect. Residuals contribute to the optimisation process and gradient flow, whereas SE blocks recalibrate channel features.

    \item The visual outcomes demonstrate enhanced accuracy along the edges of polyps and a reduced number of incorrect detections compared to U-Net, especially in complex areas with low contrast. This emphasises the importance of global context modelling.

    \item BetterNet achieves high efficiency while avoiding the computational demands associated with Transformer-based methods such as CAFE-Net, and HSNet. The compact architecture, which consists of 11.5 million parameters, allows for easy deployment on edge hardware, enabling real-time usage.
\end{enumerate}

\subsection{Clinical Translation Potential}

Accurate polyp segmentation can have high clinical utility for colonoscopy screening and surveillance:

\begin{itemize}
    \item Accurate segmentation masks provide reliable identification, monitoring, and quantification of polyps over a period of time. This can aid medical professionals in evaluating the rate of growth to determine the risk of cancer and the necessity for intervention.
    
    \item Segmentation visualisations enhance the visibility of flat or depressed polyps that may be overlooked during physical examination, hence improving the accuracy of detection.
        
    \item Segmentation enables computer-aided diagnosis by facilitating additional categorisation and risk assessment using extracted shape, size, and texture features.
        
    \item The real-time segmentation performance and minimal computational needs of BetterNet make it suitable for inclusion into smart colonoscopy devices to provide clinical decision support.
\end{itemize}

Overall, the significant performance gains, efficiency, and model interpretability demonstrated by BetterNet pave the way for translation into CADe systems for enhanced polyp screening in practice.

\subsection{Limitations and Future Work}

BetterNet focuses only on the binary segmentation of polyp versus non-polyp regions. Performing multi-class segmentation that differentiates various anatomical structures can offer more extensive contextual information.

Given our identified limitations, our upcoming focus will shift toward developing a sophisticated multi-class segmentation model that can differentiate between structures. This will offer context beyond just classifying polyps versus non-polyps. Additionally, we plan to investigate techniques for estimating uncertainty to boost the model's dependability and resilience. Collaboration with radiologists is critical in this endeavor as it allows us to access datasets covering a spectrum of anatomical variances. Our aim through these initiatives is to surpass the boundaries of our work and enhance the effectiveness of medical image segmentation for diagnosis and treatment planning.

\section{Conclusion}\label{sec:conclusion}

Precise polyp segmentation is vital in colonoscopy procedures as it facilitates accurate identification, measurement, and monitoring of polyps over time, ultimately enhancing the early detection of colorectal cancer and reducing associated mortality rates. In this work, we propose BetterNet, a novel convolutional neural network that effectively segments colonoscopy polyps by leveraging residual learning and attention mechanisms. The residual decoder combines multi-scale encoder information using skip connections and squeeze-and-excitation blocks. Ablation studies confirm the synergistic benefits of incorporating residual learning and attention mechanisms in a unified model. The lightweight design of BetterNet allows for real-time performance and is suitable for clinical translation.

Experimental results on the challenging polyp segmentation benchmarks demonstrate that BetterNet can effectively handle the difficulties posed by these datasets, such as the presence of small polyps and diverse imaging conditions. Quantitative results show that BetterNet significantly outperforms eleven existing state-of-the-art methods across multiple evaluation metrics. BetterNet proves to be highly effective in capturing fine-grained details and leveraging global context information, leading to improved segmentation accuracy. Moreover, BetterNet's lightweight design enables real-time performance, making it suitable for integration into clinical workflows and facilitating timely decision-making during colonoscopy procedures.

\section*{CRediT authorship contribution statement}
\begin{itemize}
    \item[] \textbf{O Singh}: Conceptualization, Methodology, Software, Formal analysis,  Validation, Writing - Original Draft, Visualization. 
    \item[] \textbf{S.S. Sengar}: Conceptualization, Software, Resources,  Supervision, Validation, Writing - Review \& Editing, Project administration.
\end{itemize}

\section*{Declaration of competing interest}

We affirm that no discernible competing financial interests or relationships may have unduly influenced the findings presented in this publication. Our commitment to impartiality and honesty ensures the authenticity and reliability of our research.

\section*{Data availability}

The code for the BetterNet model is open-source and available on GitHub: \url{https://github.com/itsOwen/BetterNet}. The datasets used in this study are publicly available and can be accessed from their respective sources: Kvasir-SEG (\url{https://datasets.simula.no/kvasir-seg/}), CVC-ClinicDB (\url{https://polyp.grand-challenge.org/CVCClinicDB/}), EndoScene (\url{https://figshare.com/articles/figure/Polyp_DataSet_zip/21221579}), EndoTect (\url{https://endotect.com/}), and Kvasir-Sessile (\url{https://datasets.simula.no/kvasir-seg/}).

\section*{Acknowledgements}

The work was supported by Cardiff Metropolitan University Llandaff Campus, Western Avenue, Cardiff, CF5 2YB, United Kingdom.

\bibliographystyle{cas-model2-names}
\bibliography{cas-refs}

\end{document}